\documentclass{elsart}

\usepackage{natbib}

\usepackage{amssymb}
\usepackage {graphicx}
\usepackage{placeins}
\usepackage{longtable,array,tabularx}
\begin{document}

\begin{frontmatter}



\title{NIR Imaging and Spectroscopy of AGN hosts at z $\leq$ 0.06}

\author{S. Fischer}
\author{C. Iserlohe}
\author{J. Zuther}
\author{T. Bertram}
\author{C. Straubmeier}
\author{A. Eckart}
\address{I. Physikalisches Institut, Universit\"at zu K\"oln, Z\"ulpicher Str. 77, 50937 K\"oln, Germany}

\begin{abstract}
We created a sample of nearby QSOs selected from the Hamburg/ESO survey and the V\'eron-Cetty \& V\'eron catalog, with a limiting redshift of z$\le$0.06, which consists of 63 objects.
In this contribution, we present the results of our ISAAC $Ks$-band spectroscopic and $H$ and $Ks$-band photometric observations of 9 sources of this sample.
In seven sources we find hydrogen recombination lines Pa$\alpha$ and Br$\gamma$ of which five galaxies show a broad component. 
In three sources, extended molecular hydrogen emission is detected in the 1-0S(1) line.
The stellar CO-absorption feature is only detectable in 5 objects, those sources with $J-H$ and $H-K$ colors closest to those of ordinary galaxies.
\end{abstract}

\begin{keyword}
 galaxies: active \sep  galaxies: stellar content \sep quasars: emission lines \sep quasars: absorption lines 
 \PACS 98.54.Aj \sep 98.62.Bj \sep 98.62.Lv \sep 98.62.Qz

\end{keyword}

\end{frontmatter}

\section{Introduction}
\label{intro}
  In this contribution, we present our ISAAC seeing limited NIR observations of 9 sources drawn from the Cologne Nearby QSO sample \citep[e.g.][]{bertram,fischer}.
  We performed low resolution spectroscopy with a 1'' slit in the $Ks$-band (2.2$\mu$m) which yields analysis of several diagnostic lines such as hydrogen recombination lines (Pa$\alpha$ \& Br$\gamma$) and rotational-vibrational molecular hydrogen lines.
  The stellar CO absorption bands at wavelengths $>$2.295$\mu$m allow analysis of dominating stellar spectral classes, though the absorption band is strongly affected by the rising non-stellar continuum towards the center of an AGN (see chapter \ref{spectroscopy}).\\
  In addition, we obtained $H$- (1.65$\mu$m) and $Ks$-band images of the objects what provides, with supplementary $J$-band (1.25$\mu$m) 2MASS-images, information on extinction in the galaxy and on whether the nuclear or the stellar component dominates the galaxy's radiation.

 \begin{figure}[h!]
 \begin{center}
 \includegraphics[width=5cm]{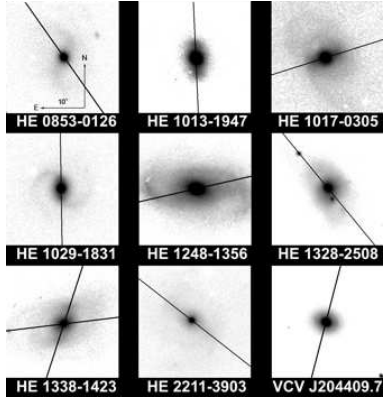}
 \caption{Our ISAAC $H$-band images of the 9 observed AGN, with slit positions of the long-slit spectroscopic observations. For HE 1248-1356, the $Ks$-band image is shown. With an average of 8s of integration time, the host galaxies are clearly resolved.
 }
 \label{fig:fig1}
 \end{center}
 \end{figure}

\section{Spectroscopy}
\label{spectroscopy}

 \begin{figure}[h!]
 \begin{center}
 \includegraphics[width=10cm]{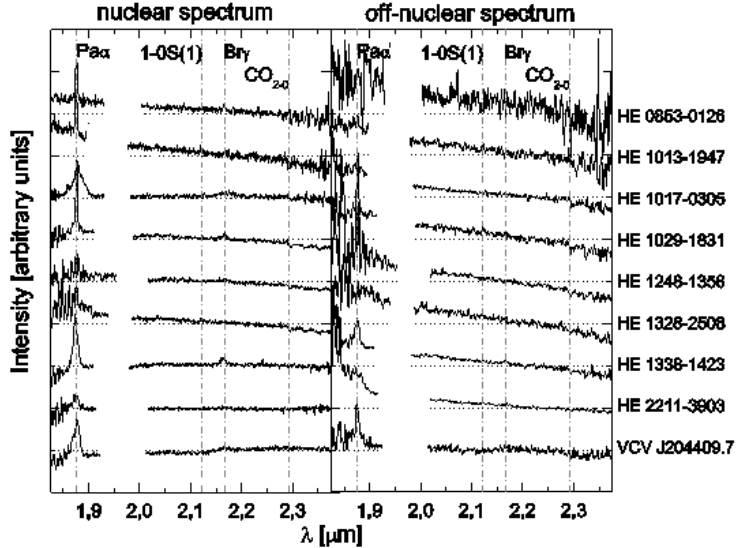}
 \caption{Extracted $Ks$-band spectra of the 9 AGN, extracted at the 
      central region (left) and at a distance of 1.5 FWHM (right, 3-point-smoothed).
      The spectra are presented in restframe, the region at $\sim$2$\mu$m is blanked out due to imperfect atmospheric correction, normalization was carried out at 2.2$\mu$m in observer frame.
      Note that the features at $>2.3\mu$m in HE 0853-0126 are not CO absorption features but noise caused by the atmosphere.
      Especially remarkable are the red continua in most central regions.
      AGN with colors close to ordinary galaxies (see Fig. \ref{fig:fig3}) correspond to spectra with bluer continuum slopes.}
 \label{fig:fig2}
 \end{center}
 \end{figure}

 	Pa$\alpha$ is detected in HE 2211-3903, HE 1013-1947, HE 0853-0126, HE 1017-0305, HE 1029-1831, HE 1338-1423 and VCV J204409.7-104324, the latter 5 show additionally Br$\gamma$ in emission.
	In the cases of HE 1013-1947, HE 1029-1831 and HE 1338-1423, the Pa$\alpha$ line shows a composition of a broad and a narrow component, while in HE 0853-0126 only a narrow component \citep[the source is classified as Sy 1 in][]{wisotzki2} and in the two remaining objects only a broad component is observed.
	For HE 2211-3903, the shape of the Pa$\alpha$ line points to more complicated kinematics.\\
	In the five cases where both hydrogen recombination lines are detected,
	the resulting extinction at the central region of two galaxies
	is low (HE 1029-1831, VCV J204409.7-104324).
	In the three other galaxies, the central region is heavily
	extincted.\\
	HE 1013-1947, HE 1029-1831, HE 1328-2508 show extended molecular hydrogen emission in the 1-0S(1) transition.
	This is a significantly lower detection rate of Sy1 galaxies with molecular hydrogen emission lines in comparison to the findings of other surveys (e.g. \citet{rodriguez} find in a sample of 22 mostly Sy1 galaxies H$_2$-emission in 90\% of their sources).
	Due to no other detected H$_2$ lines, UV-fluorescence \citep{black} can rather be excluded as excitation mechanism.\\
	In HE 1013-1947, HE 1017-0305, HE 1029-1831, HE 1248-1356 and HE 1328-2508, stellar CO absorption is detected. With the exception of HE 1328-2508, all sources show a strong increase of the CO-equivalent width with growing distance to the center. This is caused by a rising non-stellar continuum towards the nucleus. The equivalent width of the CO(2-0)-absorption in HE 1248-1356 of (11$\pm$2)\AA~away from the nucleus can most likely be associated with ongoing star formation. The other sources show equivalent widths typical of ordinary ellipticals or spirals.\\
	The continuum slopes show a correlation to the detectability of the CO-absorption. In galaxies with significant reddening, the CO-absorption is hidden in the strong non-stellar continuum.
	
 
\section{Photometry}
\label{photometry}
   	The dominating morphological class are disc dominated galaxies.
	Only one galaxy is found to be an elliptical galaxy.
	An underrepresentation of ellipticals is consistent with the results	of other samples \citep[e.g.][and references therein]{jahnke}, since most of the observed sources are lower luminosity AGN and the probability to find an underlying disc-dominated host galaxy increases	with lower luminosity nuclei.\\
	At least in HE 1017-0305 and HE 1328-2503, the appearance suggests that these objects are interacting galaxies, in the latter one a possible second nucleus is found at a distance of 3'' to the nucleus.
	This supports the theory that nuclear activity may be triggered by
	merger events.\\
	The analog trend to the continuum-slope/CO-detection can be seen when comparing the colors of the AGN to the presence of a CO-absorption feature, only those AGN with a strong stellar contribution allow for a detection of the absorption band.\\
	After subtraction of the nuclear contribution and application of a K-correction \citep[e.g. following][]{hunt}, the host galaxies show colors similar to the colors found in oridinary spiral glaxies.

 \begin{figure}[h!]
 \begin{center}
 \includegraphics[width=9cm]{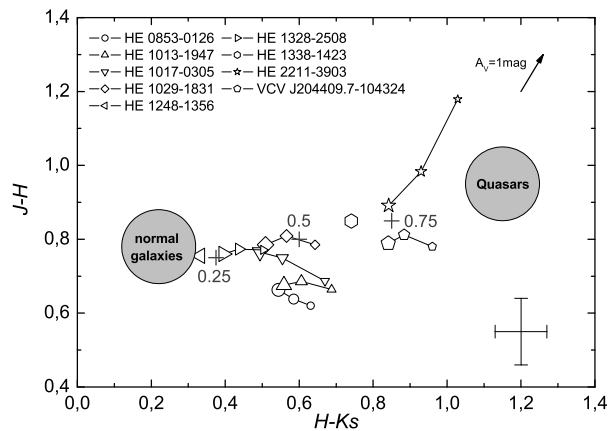}
 \caption{The NIR 2-color diagram for the 9 Sy1/NLS1 galaxies.
          Different sized data markers display flux measurements in apertures of 14'', 8'' and 3'', centered on the nucleus.
	  Additionally, the regions of normal galaxies and Quasars and the arrow specifying a visual extinction of 1${\rm ^{mag}}$ are indicated in the graph.
	  The locus of the three crosses (in grey) is based upon calculations by \citet{hyland} and refers to the radiation output of normal galaxies with a gradually increased Quasar contribution (+$_{0.25}$: 25\% nuclear radiation).}
 \label{fig:fig3}
 \end{center}
 \end{figure}





\begin{thebibliography}{}


\bibitem[Bertram et al.(2005)]{bertram}
	Bertram, T. 2005
	these proceedings

\bibitem[Black \& van Dishoeck(1987)]{black}
	Black, J. H. \& Dalgarno, A. 1981
	ApJ 249, 138

\bibitem[Fischer et al.(submitted)]{fischer}
	Fischer, S. and Iserlohe, C. and Zuther, J. and Bertram, T. and Straubmeier, C. and Sch\"odel, R. and Eckart, A. submitted to A\&A

\bibitem[Hunt et al.(1999)]{hunt}
	Hunt, L.~K. and Malkan, M.~A. and Rush, B. et al. 1999
	ApJS 125, 349                                         
	
\bibitem[Hyland \& Allen(1982)]{hyland}
	Hyland, A. R. \& Allen, D. A. 1982
	MNRAS 199, 943
	
\bibitem[Jahnke et al.(2004)]{jahnke}
	Jahnke, K. and Kuhlbrodt, B. and Wisotzki, L. 2004
	MNRAS 352, 399
	
\bibitem[Rieke \& Lebofsky(1985)]{rieke}
	Rieke, G. H. and Lebofsky, M. J. 1985
	ApJ 288, 618

\bibitem[Rodr{\'{\i}}guez-Ardila et al.(2004)]{rodriguez}
	Rodr{\'{\i}}guez-Ardila, A. et al. 2004
	A\&A 425, 457

\bibitem[Wisotzki et al.(2000)]{wisotzki2}
	Wisotzki, L. et al. 2000
	A\&A 358, 77	
	
\end{thebibliography}
\end{document}